\documentclass[11pt]{article}
\usepackage{graphicx}
\usepackage{times}
\usepackage{wrapfig}
\usepackage{setspace}
\begin{document}
\begin{center}\textit{\textbf{\large{Single-molecule imaging of protein adsorption mechanisms to surfaces}}}\\\end{center}
\begin{center}Shannon Kian Zareh and Y. M. Wang\end{center}
Department of Physics, Washington University in St. Louis, St. Louis, MO, 63130\\\\
Mailing address: 244 Compton Hall, 1 Brookings Dr., St. Louis, MO, 63130\\\\
Corresponding author:  Y. M. Wang; Address: 244 Compton Hall, 1 Brookings Dr., St. Louis, MO, 63130\\\\
Office phone: 314-935-7478

\begin{abstract}
Protein-surface interactions cause the desirable effect of controlled protein adsorption onto biodevices as well as the undesirable effect of protein fouling.  The key to controlling protein-surface adsorptions is to identify and quantify the main adsorption mechanisms: adsorptions that occur (1) while depositing a protein solution onto dry surfaces and (2) after the deposition onto wet surfaces.  Bulk measurements cannot reveal the dynamic protein adsorption pathways and thus cannot differentiate between the two adsorption mechanisms.  We imaged the interactions of single streptavidin molecules with hydrophobic fused-silica surfaces in real-time.  We observed both adsorbed proteins on surfaces and diffusing proteins near surfaces and analyzed their adsorption kinetics.  Our analysis shows that the protein solution deposition process is the primary mechanism of streptavidin adsorption onto surfaces at the sub-nanomolar to nanomolar protein concentrations.   Furthermore, we found that hydrophilic fused-silica surfaces can prevent the adsorption of streptavidin molecules. 

\end{abstract}

\begin{center}\textbf{INTRODUCTION}\end{center}

Controlled surface adsorption of proteins is important for devices such as protein-based biosensors and protein microarrays, but on the other hand, uncontrolled accumulation of proteins onto surfaces causes undesirable protein fouling (Zydney and Ho, 2003; Gray, 2004; Latour, 2005). 
In addition, when protein-surface contact is involved in a  physical process, controlling protein-surface adsorption is necessary to ensure minimal perturbation to protein concentration and characteristics in solution.  For these reasons, it is necessary to identify and quantify mechanisms responsible for protein adsorption to surfaces. 

The first step in most protein-based biological studies and applications involves introducing a protein solution to a device (e.g., a pipette tip, a transfer tube, a glass slide, etc.).  It is known that proteins can dissolve in water as well as accumulate at the air-water interface (Mackie et al., 1999; de Jongh et al., 2004; Deng et al., 2006; Yano et al., 2009). 
When these proteins encounter a device surface,  some adsorb during the protein-solution deposition process while others adsorb after the surface is wet.  These adsorptions are the result of  electrostatic,  van der Waals, and hydration interactions between proteins and surfaces (Squires et al., 2008). 
 While most studies focused on surface adsorptions of the dissolved proteins after deposition (Vasina et al., 2009), 
the effects of the deposition process on protein-surface adsorption are not clear.  

In order to precisely identify and quantify mechanisms responsible for protein-surface adsorptions,  we investigate the adsorption process of individual proteins by single molecule fluorescence imaging.  Prior studies of protein-surface interactions mainly used bulk ensemble measurements, in which the concentrations of all adsorbed proteins were measured and thus adsorptions due to different mechanisms could not be differentiated (Sapsford and Ligler, 2004; Tsapikouni and Missirlis, 2008).  
 In our study, we used Total Internal Reflection Fluorescence (TIRF) microscopy imaging method to record the interplay of a single molecule, streptavidin-Cy3 and streptavidin-Alexa555, with hydrophobic and hydrophilic fused-silica surfaces in real time.  Images of both adsorbed proteins at the surface and free 3D-diffusing proteins near the surface were captured to reveal the adsorption pathways and kinetics for both irreversible and reversible adsorptions.
\begin{center}\textbf{MATERIALS AND METHODS}\end{center}
\begin{center}\textbf{Protein Deposition}\end{center}
 Figures 1A -- 1C schematically illustrate the method of depositing a protein solution onto a surface.  Five microliters of streptavidin-Cy3 (SA1010, Invitrogen, Carlsbad, CA) or streptavidin-Alexa555 powder (S21381, Invitrogen, Eugene, OR) dissolved in 0.5X TBE buffer (pH = 8) to 0.3 nM concentration were deposited onto a fused-silica surface  (6W675-575 20C, Hoya Corporation USA, San Jose, CA) by pipette.   Streptavidin is a 52.8-kDa tetrameric protein measuring 4.5 $\times$ 4.5 $\times$ 5 nm in size (Scouten and Knecny, 1992).  It has an isoelectric point of 6.3 (Sivasankar et al., 1998) and is negatively charged in  solutions with a pH 8.  Protein concentrations of less than nanomolars were used such that images of individual adsorbed proteins on surface and 3D-diffusing proteins in solution could be resolved without overlapping.   The droplet is flattened by a coverslip, whose edges were sealed with nail polish.  Because hydrophobic surfaces are known to yield higher protein-surface adsorption affinity than hydrophilic surfaces (Israelachvili, 1992), 
the surfaces in Figs. 2, 3, and 4A were treated to be hydrophobic with a $\approx$\hspace{1mm}90$^{\circ}$ water contact angle by dipping oxygen-plasma-cleaned fused-silica chips into a 5$\%$ dichlorodimethylsilane in chloroform solution for 10 sec.  The glass coverslip was cleaned using oxygen plasma and was hydrophilic.  

 In the schematics, we showed proteins that are adsorbed on the surface, accumulated at the air-water interface, and dissolved in the solution.  Using the  aforementioned protein-solution deposition method, two interfaces are made: the initial air-water interface during the deposition process (for irreversible adsorptions), and the water-surface interface at the later stage (for reversible adsorption).
\begin{center}\textbf{Imaging Setup}\end{center}
Single-molecule imaging was performed using a Nikon Eclipse TE2000-S inverted microscope (Nikon, Melville, NY) in combination with a Nikon 100X objective (Nikon, 1.49 $N.A.$, oil immersion).   Samples were excited by a prism-type TIRF setup with a linearly polarized 532 nm laser line focused to a 40 $\mu$m $\times$ 20 $\mu$m region (568 nm line was used for streptavidin-Cy3).  The laser excitation (I70C-SPECTRUM Argon/Krypton laser, Coherent Inc., Santa Clara, CA) was pulsed with exposure times of submilliseconds and 28 ms, and the durations between images were 4 min for Fig. 2, 38 min for Fig. 3, and 30 ms for Fig. 4.  The excitation intensities were 0.3 kW/cm$^2$ to 8 kW/cm$^2$.  Images were captured by an iXon back-illuminated electron multiplying charge coupled device (EMCCD) camera (DV897ECS-BV, Andor Technology, Belfast, Northern Ireland).  An additional 2X expansion lens was placed before the EMCCD, producing a pixel size of 79 nm.  For streptavidin-Alexa555 molecules, the excitation filter was 530 nm/10 nm and the emission filter was 580 nm/60 nm; for streptavidin-Cy3 molecules, the excitation filter was 568 nm/10 nm and the emission filter was 605 nm/50 nm.
\begin{center}\textbf{Data Analysis}\end{center}
 Movies were made by synchronizing the onset of camera exposure with laser illumination.  The gain levels of the camera were adjusted such that none of the pixels of a single molecule's point spread function (PSF) reached the saturation level of the camera.  For a selected image, the intensity values of $20 \times 20$ pixels centered at the molecule were recorded.   One dimensional intensity profile of the molecule was obtained by averaging the 20 transverse pixel values at each of the 20 horizontal pixels.   The profile was then fitted to a 1D Gaussian function using the least squares curve-fitting algorithm (lsqcurvefit) provided by \textsc{MATLAB} (The Mathworks, Natick, MA):  
\begin{equation}
	f(x)=f_{0}\exp{\left[-\frac{(x-x_0)^2}{2s^2}\right]} + \langle{b}\rangle,
	\label{eqn:gauss2dfxn}
\end{equation}
where $f_0$ is the amplitude, $x_0$ is the center, $\langle{b}\rangle$ is the mean background value, and $s$ is the standard deviation (SD) of the molecule's intensity profile.  Using the SD value of a single molecule intensity profile, we determined whether the molecule was adsorbed on the surface or diffusing in the solution (see Results below).
\begin{center}\textbf{RESULTS}\end{center}
We observed three types of streptavidin molecules on or near fused-silica surfaces, categorized according to their different surface adsorption characteristics:  (1) irreversible adsorption induced by the protein deposition process, (2) reversible adsorption caused by protein interactions with wet surfaces, and (3) non-adsorbing proteins that freely diffuse near the surface.

Figure 2 shows a time series of streptavidin-Alexa555 images on or near a hydrophobic fused-silica surface, starting immediately after the deposition with time interval of 4 minutes.  The diffraction-limited dots and the larger ``blurs" are adsorbed proteins on the surface and diffusing proteins in the solution, respectively.  The long interval of 4 minutes between the images was chosen to minimize fluorophore bleaching from long illumination (Wang et al., 2005; Wang et al., 2006).  In order to determine whether a fluorescent image is an adsorbed protein or a freely diffusing protein, we measured the SD of the molecule's fluorescence intensity profile. 
If the SD value was within the diffraction limit of the apparatus ($\approx$\hspace{1mm}120 $\pm$ 20 nm), the molecule was an adsorbed protein on the surface (DeSantis et al., 2010; DeCenzo et al., 2010); 
if it was larger than 140 nm, the molecule was a diffusing protein.

The small dots whose SD values were below 140 nm that didn't change location in all  subsequent four images are denoted by yellow slanted arrows.   Because they do not change positions after the deposition, these are identified as irreversibly adsorbed proteins.  And since no additional irreversibly adsorbed proteins that were observed after the deposition, these proteins must have been adsorbed during the protein-solution deposition process.    The small dots whose SD values were below 140 nm and were observed only in one image are denoted by green vertical arrows.  Due to their momentary stay on the surface, these are reversibly bound proteins where the adsorption occurred after the deposition.  The blurred dots with SD larger than 140 nm and were only observed in one image are denoted by red horizontal arrows; these are 3D diffusing proteins near the surface.

In order to better illustrate the nature of the three types of proteins on and near the fused-silica surfaces -- (1) the irreversibly adsorbed proteins that were bound at the same locations for a long time after the sample deposition, and (2) the reversibly adsorbed proteins and (3) the 3D-diffusing proteins that change locations from one image to another, in Fig. 3 we superpositioned two images taken at different times. The first one was taken immediately after the deposition, and the second image was at 38 minutes after the deposition.  The protein dots and ``blurs" in the first image were false-colored red, and in the second image, they were false-colored green.  When the two images were superposed, the overlapped protein dots were represented by orange, and the proteins that do not overlap retain their original color.  By using this method, the orange irreversibly adsorbed proteins can be clearly differentiated from the red and green reversibly adsorbed proteins and diffusing proteins.

Streptavidin adsorption mechanisms are elucidated by analyzing Fig. 3, in which 32 molecules were irreversibly adsorbed, 2 were reversibly adsorbed, and 30 were 3D-diffusing near the surface.  These results indicate that at the molecular concentration of 0.3 nM, the main contribution to streptavidin-surface adsorption is the protein-deposition process, which accounts for  $\sim 94\%$ of the total surface-adsorbed proteins.   The protein-surface interactions after the deposition are responsible for only 2/34 $\approx$ 6$\%$ of the total adsorption.  Majority of the 3D-diffusing proteins near surfaces do not bind to the surfaces: only $\approx$\hspace{1mm}2/(30/frame$\times$6 frames=180) $\approx$ 1$\%$ proteins that encountered the surface reversibly bound to it.  We also measured various increasing streptavidin concentrations up to 5 nM (at which concentration individual protein images began to overlap, rendering adsorption studies difficult), and observed similar results as in Figs. 2 and 3.  These results indicate that the deposition process is the dominant mechanism for streptavidin adsorption to hydrophobic fused-silica surfaces for up to nM concentrations.

In order to obtain the reversible adsorption kinetics,  we used a faster frame imaging rate of 33 Hz and exposure time of 28 ms.  Figure 4A shows montage of a molecule diffusing towards, binding to, and dissociating from a surface.  Going from top to bottom of the montage, the large blurs in images 2 and 8 are the incoming molecule moving towards the surface, and the outgoing molecule leaving the surface, respectively.  The dissociation time of this molecule is $\approx$\hspace{1mm}140 ms (5 frames). About 200 reversibly bound streptavidin molecules were studied and the mean reversible binding time of streptavidin to dichlorodimethylsilane hydrophobic surfaces was $\approx$\hspace{1mm}200 ms.  

To verify that the blurred dots in Figs. 2, 3, and 4 with SD larger than 140 nm are 3D-diffusing molecules, we studied how the SDs of these molecules change with exposure time.  If they  are 3D-diffusing molecules, SD should increase with exposure time since SD is a reflection of how far a molecule diffuses during exposure.  The penetration depth of our TIRF evanescent light is $\approx$\hspace{1mm}150 nm, so for this study we chose short exposure times such that proteins  would not diffuse beyond twice of the penetration depth to ensure complete capture of the 3D-diffusing processes.  With the Brownian dynamics calculation of $\langle{x^2}\rangle = 2D_3t$, where $\langle{x^2}\rangle$ is the mean square displacement of 3D-diffusing molecules in  one direction, and $D_3 \approx$\hspace{1mm}5 $\times 10^7$ nm$^2$/s is the 3D-diffusion coefficient for streptavidin with $\approx$ 5 nm diameter,  we determined the appropriate exposure time $t$ in the sub-millisecond range so that $\sqrt{\langle{x^2}\rangle}$ is less than 300 nm.

Figures 5A, 5B, and 5C show representative 3D-diffusing streptavidin-Cy3 molecules' intensity profiles near a hydrophilic fused-silica surface with increase exposure times of 0.3, 0.7, and 1 ms, respectively.   It is readily seen that the width of the molecules' intensity profiles increases with exposure time.  Figure 5D shows the SD distributions for the three exposure times: the mean SD values are larger than 140 nm and increasing with exposure time.  This confirms that the observed molecules with SD larger than 140 nm are indeed 3D-diffusing streptavidin-Cy3 molecules near the surface.

 So far we showed that for a ``sticky" hydrophobic surface, the dominating mechanism responsible for streptavidin fused-silica surface adsorption at sub-nanomolar and nanomolar concentrations is the deposition process.  For hydrophilic surfaces that are believed to be less ``sticky" to  the same proteins, how does the protein-surface adsorption change?  Fused-silica chips were made hydrophilic by performing oxygen plasma cleaning for 2 minutes.   We found complete elimination of irreversible and reversible streptavidin-surface adsorptions for several hours of observation time.   Figure 4B shows that there were only 3D-diffusion proteins near the hydrophilic surface and no adsorbed proteins.  This result indicates that  the surface treatment by oxygen plasma cleaning can prevent streptavidin fouling on fused-silica surfaces.  
\begin{center}\textbf{DISCUSSION}\end{center}
\begin{center}\textbf{Deposition-process-associated irreversible adsorptions}\end{center}
What is the possible mechanism that causes the irreversible protein adsorption during deposition?   Assuming that the surface is chemically stable from the moment water touches the surface,  during the deposition, a dissolved protein should have the same binding affinity to the surface as after the deposition.  However, during the deposition process, air is an additional component in the protein-surface interface and this may result in a different protein-surface binding affinity and consequently, irreversible adsorption for \textit{proteins at the air-water interface}. 

To test this hypothesis, we transferred protein solutions to a wet surface and imaged in real time.   Since there was no air-water interface in this setup, only dissolved protein in the solution can be imaged, therefore no irreversibly adsorbed proteins should be observed.  We indeed observed only reversibly adsorbed proteins on surfaces and diffusing proteins in solution.  No irreversibly adsorbed proteins were seen.  The result indicates that the proteins at the air-water-interface are the irreversibly adsorbed proteins, and this adsorption occurred only during the deposition process.  

 We further investigated the ratio of protein concentrations at air-water interface and in bulk by imaging air-water-interface proteins in a droplet on glass, using the method described by Deng et al., 2006.  We found the concentrations were comparable.  This observation is consistent with our results in Fig. 3 that half of the proteins imaged in one snapshot were irreversibly adsorbed on the surface, and thus supports the notion that the air-water interface proteins are the irreversibly adsorbed ones during the deposition process.
\begin{center}\textbf{Reversible adsorptions}\end{center}
The reversible adsorptions that occurred after the deposition are due to genuine interaction of dissolved streptavidin with surface chemical groups.  These interactions include hydrophobic, ionic, and van der Waals interactions (Sequires et al., 2008; Heinz et al., 2009).   Streptavidin is negatively charged and hydrophilic in a buffer at pH 8 (van Oss et al., 2003).  The fused-silica surface groups  can be dimethylsilane or silanol, depending on whether the surface is hydrophobic or hydrophilic.  Our observation of reversibly adsorbed proteins  on hydrophobic surfaces but not on hydrophilic ones indicates that the net binding affinity between streptavidin and the surface groups is strong enough for binding to occur for the hydrophobic surfaces, but not  for the hydrophilic surfaces.  

\begin{center}\textbf{Other hydrophobic surfaces and proteins}\end{center}
To determine the role hydrophobicity plays in streptavidin-surface adsorptions, we investigated streptavidin interaction with three differently treated hydrophobic surfaces:  RainX (SOPUS Products, Houston, TX), lab detergent (Versa-Clean, 04-342, Fisher Scientific, Pittsburgh, PA), and 0.1 wt$\%$ solution of dodecyltrichlorosilane in hexane.  The contact angles for these surfaces were $\approx$\hspace{1mm}90$^{\circ}$.   The adsorption results were the same as for the dichlorodimethysilane treated surfaces: we observed both irreversible and reversible adsorptions, with the irreversible adsorptions outnumbering the  reversible adsorptions.   We also changed the degree of hydrophobicity by tuning the ratio of dichlorodimethylsilane to chloroform, and hence the contact angles from approximately 30$^{\circ}$ to 90$^{\circ}$.  With decreasing hydrophobicity, we observed less irreversible and reversible adsorptions .   These observations indicate that hydrophobicity of a fused-silica surface can effectively dictate adsorption of streptavidin (or proteins).  Further supports come from our studies of fused-silica surface adsorptions of green fluorescent proteins and Lactose repressor proteins; similar results to streptavidin were obtained.

\begin{center}\textbf{Competition for surface binding between protein and water molecules}\end{center}
In addition to proteins, water molecules also have binding affinity to surfaces, and they compete with streptavidin for surface binding.  This competition may also contribute to the observed streptavidin adsorptions.

In Figs. 1D and 1E we  sketch possible competitions between a protein molecule and water molecules for binding to hydrophobic and hydrophilic surfaces, respectively.  Since streptavidin is hydrophilic at pH 8, when both streptavidin and water approach a hydrophobic surface, water tends to avoid the surface more than streptavidin, leading to an effectively increased protein exposure to the surface and consequently increased adsorption (Fig. 1D).  For the hydrophilic surface, water molecules are  more tightly bound to the surface, leading to an effectively decreased streptavidin adsorption (Fig. 1E).

\begin{center}\textbf{Deposition variations}\end{center}
Variations to procedures in transferring proteins to devices  include changing the fluid flow speed.  We varied the pipetting speed when depositing proteins onto dry hydrophobic and hydrophilic surfaces, varying the fluid flow speed by at least 10-fold.  No difference in irreversible and reversible adsorption characteristics was observed.
\begin{center}\textbf{CONCLUSION}\end{center}

In summary, single-molecule real-time imaging of protein-surface interactions provides an  invaluable tool for elucidating adsorption mechanisms and obtaining adsorption kinetics.  We have shown that irreversible and reversible adsorptions are highly process dependent at the sub-nanomolar to nanomolar concentrations.   Our results indicate that in addition to regulating post-deposition protein-surface interactions, the deposition process must be taken into consideration in the design and interpretation of protein-surface adsorption studies.  The observation that the surface adsorption of streptavidin can be  eliminated or reduced with hydrophilic surface treatment may have important implications for prevention of protein fouling in biomedical devices.

\begin{center}\textbf{References}\end{center} 
DeCenzo SH, DeSantis MC, Wang YM. 2010. Single-image separation measurements of two unresolved fluorophores. Opt Express 18(16):16628-16639. \\\\
de Jongh HHJ, Kosters HA, Kudryashova E, Meinders MBJ, Trofimova D, Wierenga PA, 2004. Protein Adsorption at Air-Water Interfaces: A Combination of Details. Biopolymers 74:131-135. \\\\
Deng Y, Zhu YY, Kienlen T, Guo A. 2006. Transport at the air/water interface is the reason 
for rings in proteins microarrays. J AM Chem Soc Comm 128:2768-2769. \\\\
DeSantis MC, DeCenzo SH, Li JL, Wang YM. 2010. Precision analysis for standard deviation measurements of immobile single fluorescent molecule images. Opt Express 18(6):6563-6576. \\\\
Gray JJ. 2004. The interaction of proteins with solid surfaces. Curr Opin Struct 
Biol 14:110-115. \\\\
Heinz H, Farmer BL, Pandey RB, Slocik JM, Patnaik SS, Pachter R, Naik RR, 2009. Nature of Molecular Interactions of Peptides with Gold, Palladium, and Pd-Au Bimetal Surfaces in Aqueous Solution. J Am Chem Soc 131:9704-9714. \\\\
Israelachvili JN. 1992. Intermolecular and surface forces. 2nd edn. Academic Press.\\\\
Latour RA. 2005.  Biomaterials: Protein-Surface Interactions. New York: Taylor $\&$ Francis. 1p.\\\\
Mackie AR, Gunning AP, Wilde PJ, Morris VJ, 1999. Orogenic Displacement of Protein from the Air/Water
Interface by Competitive Adsorption. J Colloid Interface Sci
210:157-166. \\\\
Sapsford KE, Ligler FS. 2004. Real-time analysis of protein adsorption to a variety of thin films. Biosens and Bioelectron 19:1045-1055.\\\\ 
Scouten WH, Konecny P. 1992.  Reversible immobilization of antibodies on magnetic beads.  Anal. Biochem 205:313-318.\\\\
Sivasankar S, Subramaniam S, Leckband D, 1998. Direct molecular level measurements of the electrostatic properties of a protein surface. Proc Natl Acad Sci 95:12961-12966.\\\\ 
Squires TM, Messinger RJ, Manalis SR. 2008. Making it stick: convection, reaction and 
diffusion in surface-based biosensors. Nat Biotechnol 26(4):417-426.\\\\
Tsapikouni TS, Missirlis YF. 2008. Protein-material interactions: From micro-to-nano 
scale. Mater Sci Eng B 152:2-7.\\\\ 
Vasina EN, Paszek E, Nicolau DV, Nicolau DV. 2009. The BAD project: data mining, 
database and prediction of protein adsorption on surfaces. Lab Chip 9:891-900.\\\\
van Oss CJ, Giese RF, Bronson PM, Docoslis A, Edwards P, Ruyechan WT, 2003. Macroscopic-scale surface properties of streptavidin and their
influence on aspecific interactions between streptavidin and
dissolved biopolymers. Colloids Surf., B 30:25-36.\\\\
Wang YM, Austin RH, Cox EC. 2006. Single molecule measurements of repressor protein 
1D diffusion on DNA. Phys Rev Lett 97:048302(4).\\\\
Wang YM, Tegenfeldt J, Reisner W, Riehn R, Guan XJ, Guo L, Golding I, Cox EC, 
Sturm J, Austin RH. 2005. Single-molecule studies of repressor-DNA interactions show 
long-range interactions. Proc Nat Acad Sci USA 102:9796-9801.\\\\
Yano YF, Uruga T, Tanida H, Toyokawa H, Terada Y, Takagaki M, Yamada H, 2009. Driving Force Behind Adsorption-Induced Protein Unfolding: A
Time-Resolved X-ray Reflectivity Study on Lysozyme Adsorbed at an
Air/Water Interface. Langmuir 25:32-35.\\\\
Zydney AL, Ho CC. 2003. Effect of membrane morphology on system capacity during normal flow microfiltration.  Biotechnol. Bioeng 83:537-543.

\begin{center}\textbf{Figure Legends}\end{center} 
\textbf{Figure 1.}
(A) -- (C) Schematic of the protein (orange) deposition process.  (D) and (E), protein and water (blue) compete for binding to hydrophobic and hydrophilic surfaces, respectively.\\\\
\textbf{Figure 2.}
(A) -- (D), time series of streptavidin-Alexa555 adsorbed on and diffusing near a hydrophobic fused-silica surface.  The first image was acquired right after the deposition, and other images were 4 minutes apart.  Yellow slanted arrows indicate irreversibly bound proteins due to the deposition process (denoted in B); green vertical arrows indicate reversibly bound proteins after the deposition (A -- D); and red horizontal arrows indicate 3D-diffusing proteins (A -- D). The scale bar is 1 $\mu$m.\\\\
\textbf{Figure 3.}
Superposed TIRF images of the streptavidin-Cy3 molecules immediately after deposition (false-colored red) and 38 minutes after the deposition (false-colored green).  When the images overlap, the red and green dots yield orange dots (denoted by orange solid arrows).  The single-colored red and green molecules with SD less than 140 nm (solid green and red arrows) are reversibly adsorbed proteins after the deposition.  The large green and red blurs are 3D-diffusing molecules near the surface (dashed red and green arrows).   The scale bar is 1 $\mu$m.\\\\
\textbf{Figure 4.}
(A) Montage of a reversibly bound molecule diffusing towards, binding to, and dissociating from the surface, respectively from top to bottom.  (B) 3D-diffusing proteins near a fused-silica hydrophilic surface.  Note that there are no adsorbed proteins on the surface.  Scale bars are 1 $\mu$m.\\\\
\textbf{Figure 5.}
Images of representative single 3D-diffusing molecules with exposure times of (A) 0.3 ms, (B) 0.7 ms, and (C) 1 ms.  The width of the molecules increases with exposure time and the 1D fit SD values are 135 nm (A), 180 nm (B), and 204 nm (C), respectively.  (D) SD distribution of the diffusing molecules' intensity profiles for exposure times of 0.3 ms (red), 0.7 ms (blue), and 1 ms (yellow).  The SD values are 139.5 $\pm$ 3.6 nm (mean $\pm$ standard error of the mean), 173.3 nm $\pm$ 4.2 nm, and 194.5 $\pm$ 5.2 nm, respectively.  The scale bar is 1 $\mu$m.

\begin{figure}[htbp]
\begin{center}
\centerline {
\includegraphics[width=3.4in]{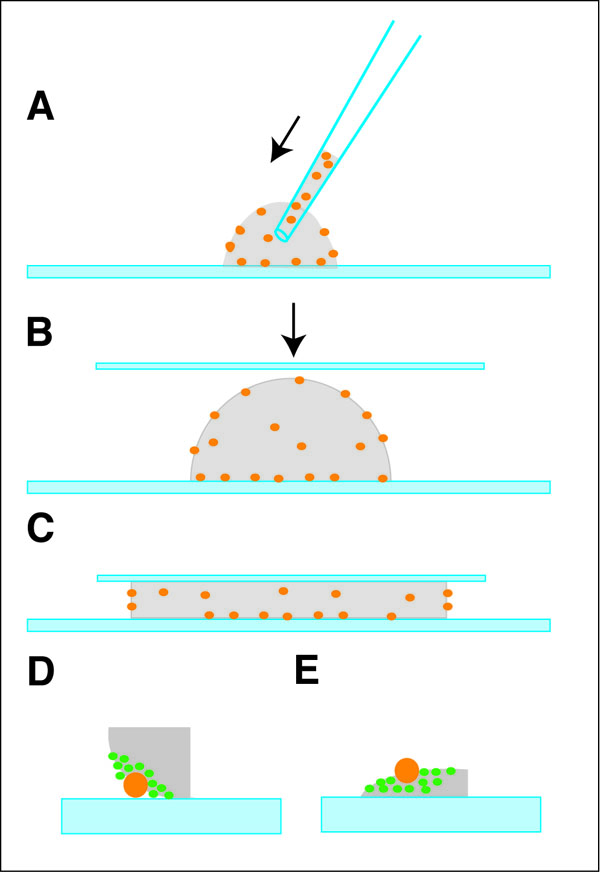}}
\caption{}
\label{Fig1}
\end{center}
\end{figure}

\begin{figure}[htbp]
\begin{center}
\centerline {
\includegraphics[width=5.4in]{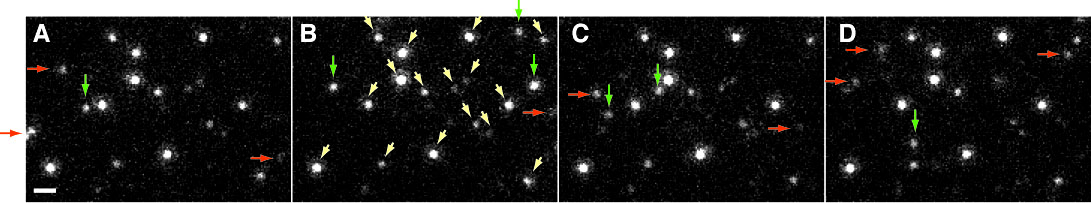}}
\caption{}
\label{Fig2}
\end{center}
\end{figure}

\begin{figure}[htbp]
\begin{center}
\centerline {
\includegraphics[width=3.4in]{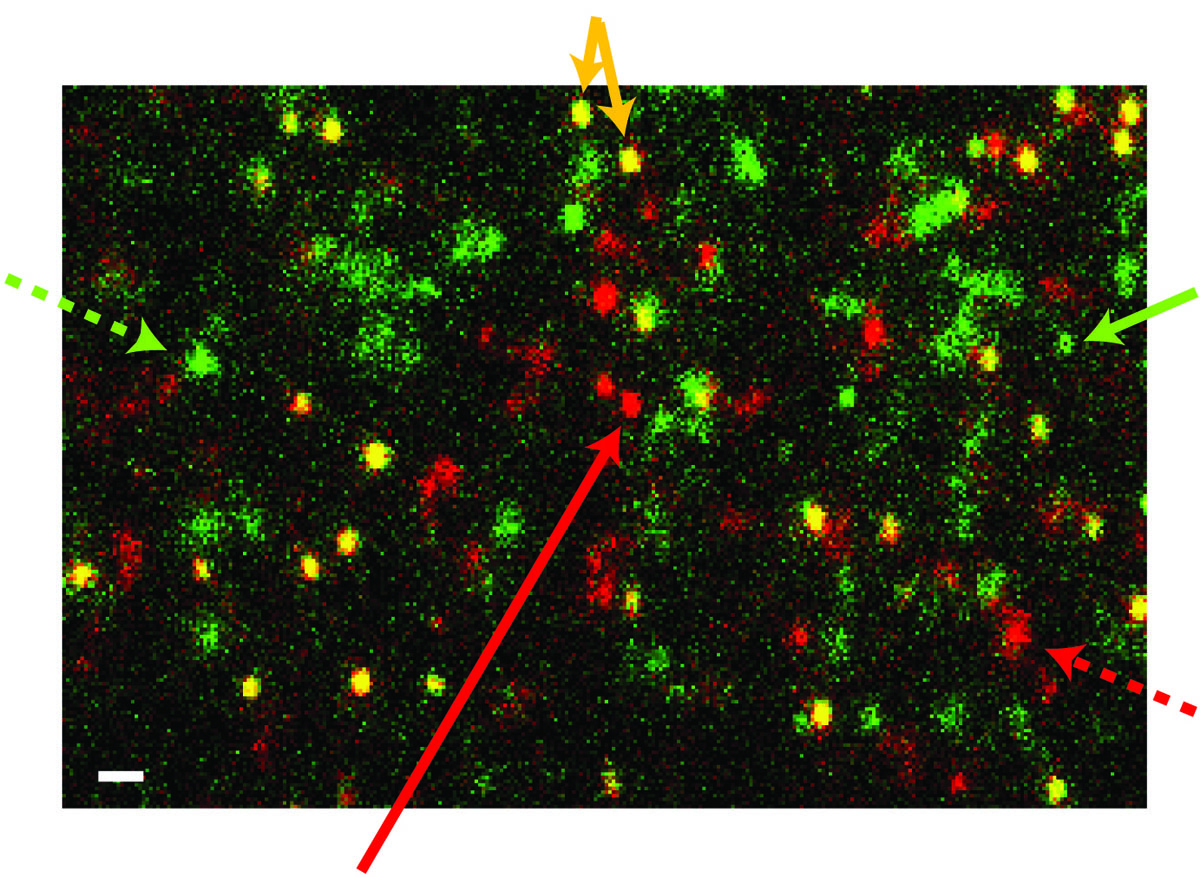}}
\caption{}
\label{Fig3}
\end{center}
\end{figure}

\begin{figure}[htbp]
\begin{center}
\centerline {
\includegraphics[width=3.4in]{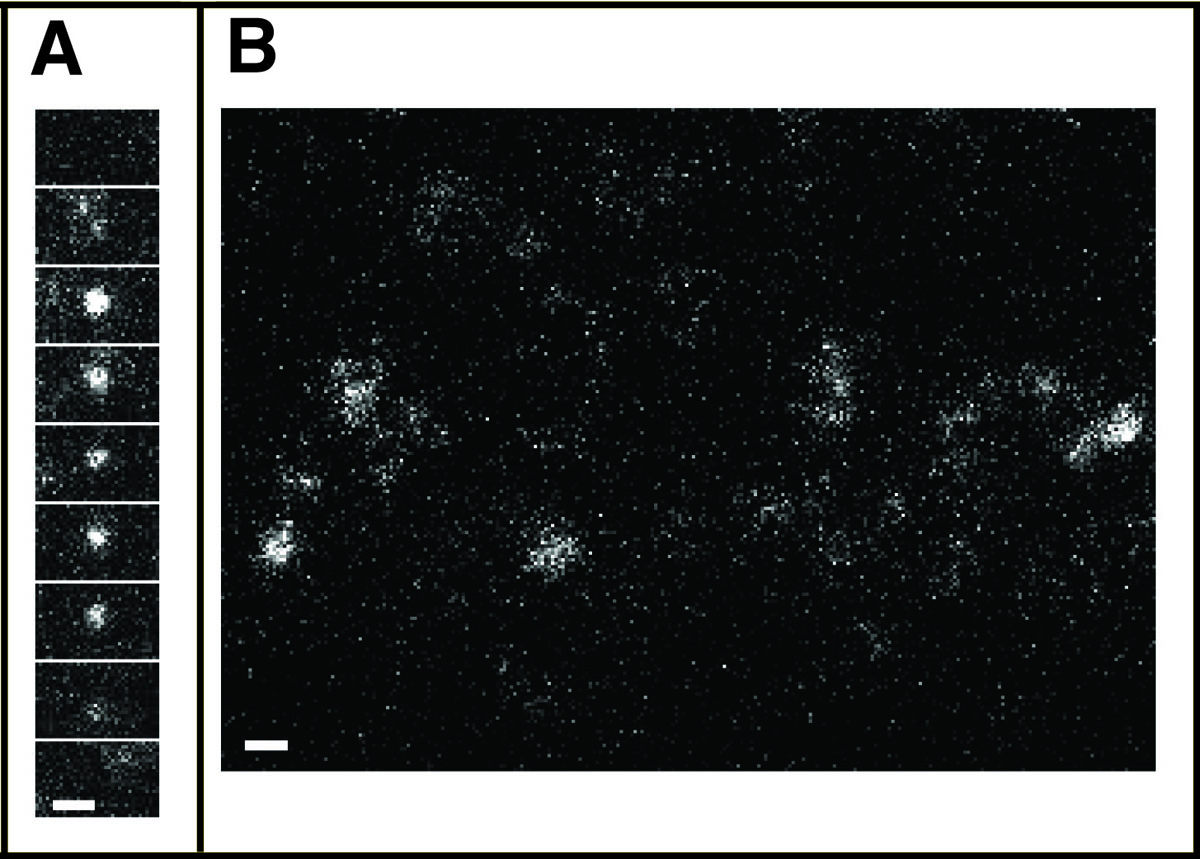}}
\caption{}
\label{Fig4}
\end{center}
\end{figure}

\begin{figure}[htbp]
\begin{center}
\centerline {
\includegraphics[width=3.4in]{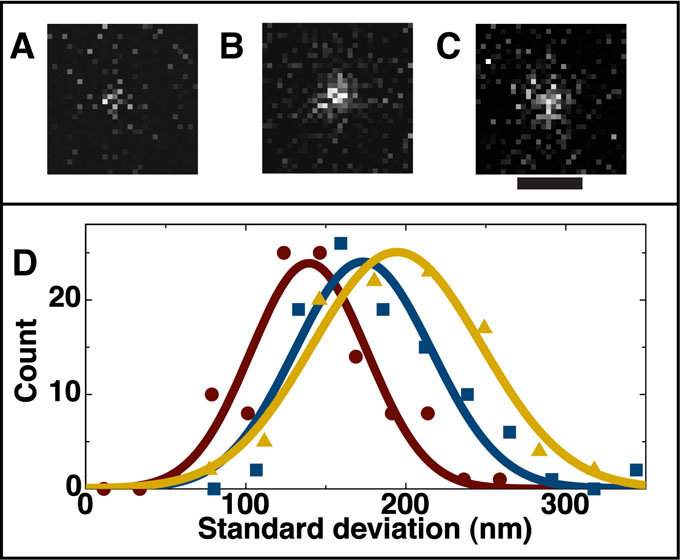}}
\caption{}
\label{Fig5}
\end{center}
\end{figure}

\end{document}